\documentclass{aa}
\usepackage{graphicx}
\usepackage{natbib}
\usepackage{amssymb,amsmath}
\bibpunct{(}{)}{;}{a}{}{,}
\usepackage[varg]{txfonts}
\usepackage{subfigure}
\usepackage{epstopdf}
\usepackage{float}
\usepackage{longtable}
\usepackage{hyperref}
\usepackage{color}
\usepackage{orcidlink}
\begin{document}

   \title{Revisiting the energy distribution and formation rate of CHIME fast radio bursts}

   \subtitle{}

\author{K. J. Zhang\textsuperscript{\orcidlink{0000-0001-6589-4795}}\inst{1,6}
         \and
        Z. B. Zhang\textsuperscript{\orcidlink{0000-0003-4859-682X}} \inst{1,*}
                  \and         
        A. E. Rodin\textsuperscript{\orcidlink{ 0000-0003-2649-7337}} \inst{2}
         \and
         V. A. Fedorova\textsuperscript{\orcidlink{0000-0002-5545-5024}}\inst{2}
     \and
     Y. F. Huang\textsuperscript{\orcidlink{0000-0001-7199-2906}}\inst{3} 
              \and
     D. Li\textsuperscript{\orcidlink{0000-0003-3010-7661}}\inst{4,5}
      \and
     X. F. Dong\inst{3}
       \and   
     P. Wang\textsuperscript{\orcidlink{ 0000-0002-3386-7159}}\inst{5}
  \and
  Q. M. Li\inst{1}
  \and
 C. Du\inst{3} 
  \and
 F. Xu\inst{3} 
   \and
 C. T. Hao\inst{1}
}
\institute{Department of Physics, Guizhou University, Guiyang 550025, P. R. China.            \email{z-b-zhang@163.com}\label{inst1}
     \and
 P. N. Lebedev Physical Institute of the Russian Academy of Sciences, Leninsky prospekt, 53, Moscow, 119991, Russia\label{inst2}
 \and
School of Astronomy and Space Science, Nanjing University, Nanjing 210023, P. R. China\label{inst3}
\and
New Cornerstone Science Laboratory, Department of Astronomy, Tsinghua University, Beijing 100084,  P. R. China\label{inst4}
\and
National Astronomical Observatories, Chinese Academy of Sciences, A20 Datun Road, Beijing 100101, P. R. China\label{inst5}
\and
South-Western Institute for Astronomy Research, Yunnan University, Kunming, Yunnan 650504, P. R. China\label{inst6}
             }

   \date{Received; accepted:accepted for publication in $A\&A$}

\abstract{Based on the first CHIME/FRB catalogue, three volume-limited samples of fast radio bursts (FRBs) are built, with samples 1, 2, and 3 corresponding to a fluence cut of 5, 3, and 1, respectively. The Lynden-Bell's c$^-$ method was applied to study their energy function and event rate evolution with redshift ($z$). Using the non-parametric Kendall's $\tau$ statistics, it is found that the FRB energy ($E$) strongly evolves with redshift as $E(z)\varpropto(1+z)^{1.24}$ for sample 1, $E(z)\varpropto(1+z)^{0.98}$ for sample 2, and $E(z)\varpropto(1+z)^{1.99}$ for sample 3. After removing the redshift dependence, the local energy distributions of the three samples can be well described by a broken power-law form with a broken energy of $\sim10^{40} \rm erg$. Meanwhile, the redshift distributions of samples 1 and 2 are identical but different from that of sample 3. Interestingly, we find that the event rates of samples 1 and 2 are independent of redshift, and sample 3 decreases as a single power-law form with an index of -2.41. The local event rates of the three samples of CHIME FRBs are found to be consistently close to $\sim 10^4\rm{\,Gpc^{-3}yr^{-1}}$, which is comparable with some previous estimates. In addition, we notice that the event rate of sample 3 FRBs with lower energies not only exceeds the star formation rate at the lower redshifts but also always declines with the increase in redshift. We suggest that the excess of FRB rates compared with the star formation rate at low redshift mainly results from the low-energy FRBs that could originate in the older stellar populations.}

   \authorrunning{ }            
   \titlerunning{ }  
   \maketitle

%
\section{Introduction} \label{sec:intro}         

A fast radio burst (FRB) is a bright flash of radio waves that has a very short timescale, from a few milliseconds to several seconds \citep[e.g.][]{Petroff2019review,li2021long}, and a large dispersion measure (DM), showing an apparent extragalactic origin. Over the past decade, more and more researchers have been motivated to seek out FRBs' physical origins. Unfortunately, the progenitors and central engines of FRBs are still unclear in nature. Many authors have proposed that they may be associated with star formation regions inside galaxies and can be explained by different progenitors such as black holes (BHs), magnetars, supernovae (SNe), and neutron stars (NSs).\citep[e.g.][]{platts2019living,zhang2020physical,Xiao2021frbrev,petroff2022fast}\footnote{See \url{https://frbtheorycat.org}}. The accurate localization of FRBs in their host galaxies can offer good clues about their origins. The host galaxies of roughly two dozen FRBs have been successfully identified\footnote{See  \url{https://github.com/FRBs/FRBhostpage}} \citep[][reference therein]{heintz2020host}. For instance, the detection of a bright FRB 20200428 originating in the Galactic magnetar SGR 1935+2154 provides an important link between FRBs and magnetars \citep{andersen2020bright,bochenek2020fast,lin2020no,mereghetti2020integral,li2021hxmt,ridnaia2021peculiar,tavani2021x,wada2023expanding}. Thus, the magnetar model is one of the most likely explanations for FRBs as the magnetar-based models can reproduce multiple FRB characteristics \citep{zhang2020physical}. Importantly, magnetars can be formed in both young (via core-collapse SNe) and old (namely, white dwarf (WD) -NS coalescence, accretion-induced collapse, and other compact binary coalescences) star populations. 

Good localization can also help to identify FRB progenitors according to distinct galactic environments \citep{kremer2021dynamical,2023Parti...6..451P}. It has been found that most well-localized FRBs
such as FRB 20121102A \citep{tendulkar2017host}, FRB 20190520B \citep{niu2022repeating}, and FRB 20190303A \citep{michilli2023subarcminute} are generally generated from active or quasi-star-forming galaxies \citep[eg.,][]{tendulkar2017host,bannister2019single,prochaska2019low,marcote2020repeating,fong2021chronicling,ravi2022host,niu2022repeating,bhandari2023nonrepeating,2023ApJ...954...80G,2024ApJ...967...29L}. In this situation, it will be expected that young star populations should be favored as the origin of FRBs.
However, there are some well-localized FRBs whose host galaxies are quiescent instead of star-forming. For example, optical observations reveal that FRB 20201124A is situated in an inter-arm region of a barred-spiral galaxy, in which the galactic environment does not comply with the young star populations \citep{2022Natur.609..685X}. In addition, FRB 20220509G has been recorded as localizing in a quiescent galaxy \citep{2024ApJ...967...29L}. Furthermore, the host galaxy location of FRB 20180916B was found to deviate from the star-forming region \citep{tendulkar202160}. Therefore, the explanation of this phenomenon is that at least some FRBs are associated with old star populations, namely NSs and BHs \citep[see, e.g.,][]{mottez2014radio,zhang2017cosmic,deng2018fast,zhang2022chime}. In particular, FRB 20200120E has been identified as being within an old globular cluster with an age of 9 Gyr in the extragalaxy M81 \citep{bhardwaj2021nearby,kirsten2022repeating}. As a whole, FRB host galaxies have two kinds of environments: the star formation region \citep[eg.,][]{tendulkar2017host,bassa2017frb,bannister2019single,tendulkar202160,niu2022repeating} and the star quiescent region \citep[eg.,][]{bhardwaj2021nearby,kirsten2022repeating}. This shows that FRBs might originate in multiple galactic environment channels. However, a very limited number of well-localized FRBs is inadequate to reveal the FRB's origins. 

With the increase in FRB numbers, a statisical study of luminosity ($L$) or energy ($E$) distributions is an alternative solution to disclose their progenitors. In other words, the $L$ or $E$ evolution with redshift, or the  formation rate (or co-moving density) evolution, has been found to be an effective way to constrain the FRB origin models \citep[e.g.,][]{locatelli2019luminosity,hashimoto2020no,arcus2021fast,james2022fast,zhang2022chime}. Although the  underlying physics of the FRBs still remains uncertain, the total energy output should be linked to the FRB progenitors. On the other hand, the progenitors of some FRBs have been found to connect with the star-forming activity, which motivates us to explore the FRB origins by describing how the $L$ or $E$ function and formation rate evolve with redshift. By contrast, the isotropic energy is reliably measured and the luminosity is usually uncertain, since the intrinsic pulse width of most FRBs is unknown \citep{macquart2018fast,ravi2019observed,pleunis2021fast,2025ApJ...980..114W} and the pulse widths of FRB 20220912A are  frequency-dependent \citep{2023ARep...67..970F}. Consequently, we analyze the energies instead of the luminosities of FRBs in this work. The FRB energy function and redshift evolution may provide an even more representative clue to constrain the possible FRB progenitors. Recently, \cite{hashimoto2022energy} found that if FRBs result from younger star populations tracking the cosmic star formation history, the FRB formation rate should increase with increasing redshift up to z $\sim$ 2. They also speculated that the FRB formation rate may decrease towards higher redshifts if the origin of FRBs is related to old populations. This will help us to constrain the  progenitor models and plan the most adequate follow-up observations, and will provide further clues to elucidate the origin of FRBs.

To estimate the formation rate of FRBs, several statistical methods, such as parametric methods including the direct fitting technique, Bayesian statistics \citep[eg.,][]{luo2018normalized,luo2020frb,tang2023inferring}, and the likelihood estimation \citep{marshall1983analysis}, and  non-parametric methods such as the $V_{max}$ method \citep{hashimoto2020no,hashimoto2020luminosity,hashimoto2022energy}, the $V/V_{max}$ method \citep{locatelli2019luminosity}, and the popular Lynden-Bell's c$^-$ method, \cite[e.g.,][]{efron1999nonparametric,petrosian2015cosmological,yu2015unexpectedly,deng2019energy,2022MNRAS.513.1078D}, have been applied in the past (see also \cite{2023ApJ...958...37D} for a short review).  
For this purpose, the isotropic energy, $E$, or luminosity, $L$, is usually used to plot against the redshift, $z$, where $E$ or $L$ increases towards higher redshifts. This is a common observational effect seen for any extragalactic source, including galaxies, quasars, gamma-ray bursts, and FRBs. At lower redshifts, the survey volume is not sufficient to detect rare bright sources but abundant faint populations are detectable, whereas at higher redshifts, the survey volume is large enough to detect the rare bright population but the sensitivity is not enough to detect faint population. This effect makes an apparent correlation between the $E/L$ and $z$ \citep{locatelli2019luminosity,hashimoto2020no,james2022fast} and mainly results from the instrument threshold truncation \cite[see however][]{2022MNRAS.513.1078D}, which can be examined by the $V_{max}$ method  \citep{hashimoto2020no,hashimoto2020luminosity,hashimoto2022energy,2023Univ....9..251Z}. This instrumental effect has been taken into account in previous works \citep[see e.g.][]{locatelli2019luminosity,hashimoto2020no,james2022fast}. To deal with this effect carefully, \cite{deng2019energy} and \cite{zhang2022chime} conducted some focused analyses. To mitigate instrumental selection effects between different telescopes, it is a typical practice to use a large and homogeneous sample collected by a single telescope. Nevertheless, CHIME missed a significant number of FRBs with fluences near the threshold, which causes these populations to be absent in the first CHIME catalogue. To make the analysis free of this observational bias, a selection function \cite[e.g.][]{hashimoto2022energy,2023ApJ...944..105S} has to be utilized during the data analyses. In other words, a volume-limited sample (see Sect. \ref{sec2.2:data} for details) should be adopted for the estimation of event rates. We then use the Kendall's $\tau$ statistics proposed by \citet{EP1992simple,efron1999nonparametric} to examine the correlation between energy and redshift. Then, we apply the Lynden-Bell's c$^-$ methods to obtain the  distributions of energy de-evolved with redshift and the formation rate history.  
  
This paper is organized as follows. Section \ref{sec2: data} describes the data reduction process. Section \ref{sec:method} presents the Lynden-Bell's c$^-$ method and the non-parametric Kendall's $\tau$ statistics. We derive the FRB energy function in Sect. \ref{energyfun} . The redshift distribution and formation rate are shown in Sect. \ref{sec:zdisandFm}. In Sect. \ref{sec:localeventrate}, we give the local event rate of the sample of CHIME FRBs. Finally, we end with conclusion and discussions in Sect. \ref{sec:concl}. Throughout the paper, a flat $\Lambda$CDM universe with $H_{0}=67.74\,\mathrm{km} \mathrm{s}^{-1} \mathrm{Mpc}^{-1}$, $\Omega_{\Lambda}=0.69$, and $\Omega_{m}=0.31$ \citep{Planck2016} is assumed.
\section{Data preparation}\label{sec2: data}
\subsection{Dispersion measure}\label{sec2.1:DM}
In general, the observed DM of a FRB is predominantly contributed to by electrons in the line of sight and can be roughly written as
        \begin{equation}
                DM=DM_{MW}+DM_{halo}+DM_{IGM}+\frac{DM_{host}+DM_{src}}{1+z},
                \label{eq:DM-1}
        \end{equation}
or
\begin{equation}
        DM=DM_{MW}+DM_{E},
                \label{eq:DM-2}
\end{equation}
where the subscripts $MW$, $halo$, $IGM$, $host$,  $src$, and $E$ denote the DM contributions of the Milky Way (MW), the halo of the MW, the intergalactic medium, the FRB host galaxy, the circum-source environment, and the extrogalactic dispersion, respectively. The $DM_{MW}$ can be estimated by the NE2001 model \citep{2002astro.ph..7156C} or YMW16 model \citep{Yao2017}. The DM portion of our Galactic halo is about $DM_{halo}=10-80$ pc cm$^{-3}$ \citep{2019MNRAS.485..648P} or even lower than $10$ pc cm$^{-3}$ \citep{2020MNRAS.496L.106K}. The IGM portion of DM is related to the cosmological redshift by \citep{2014ApJ...783L..35D,zhang2018frb}
        \begin{equation}
                DM_{IGM}=\frac{3cH_0\Omega_bf_{IGM}}{8\pi Gm_p}\int_{0}^{z} \frac{\chi(z)(1+z)}{[\Omega_m(1+z)^3+\Omega_{\lambda}]^{1/2}}\mathrm{d}z, 
                        \label{eq:DMIGM}
        \end{equation}
        in which all parameters can be referred to the literature therein. The $DM_{host}$ similar to $DM_{MW}$ is also largely uncertain and could have a large range of $DM_{host}\approx 55-275$ pc cm$^{-3}$ \citep{tendulkar2017host}. The DM contribution from the ambient environment of FRB source is unknown too. Following \citet{zhang2018frb}, we also take $DM_{src}\approx 0$ pc cm$^{-3}$ as a reliable assumption, since the size of the circum-burst atmosphere should be much smaller than the sizes of the galaxies and the intergalactic medium. In addition, two typical values of $DM_{halo}\approx 50$ pc cm$^{-3}$ and  $DM_{host}\approx 100$ pc cm$^{-3}$ have been reasonably assumed to select FRB samples as follows. Thus, Eqs. (\ref{eq:DM-1}) and (\ref{eq:DM-2}) can be combined into
                \begin{equation}
        DM_E=DM-DM_{MW}\approx 50     \mathrm{ pc\ cm^{-3}}+DM_{IGM}+\frac{100 \mathrm{ pc\ cm^{-3}}}{1+z},
                \end{equation}  
        which can be applied to estimate the redshift of a FRB under the condition that the value of $DM_{IGM}$ in eq. (\ref{eq:DMIGM}) is positive. For simplicity, people often utilize $z \sim DM_{E}/[855\ \mathrm{ pc\ cm^{-3}}]$ to estimate the redshift that is solely an upper limit \cite[e.g.][]{zhang2018frb}. The relatively precise redshift should directly depend on $DM_{IGM}$; namely, $z \sim DM_{IGM}/[855\ \mathrm{ pc\ cm^{-3}}]$. The statistical errors of $z$ can be evaluated through an assumption of 40\% variation in the conversion factor of 855 \citep{zhang2018frb}.

\subsection{Sample selection} \label{sec2.2:data}

\begin{figure*}
        \includegraphics[width=0.5\textwidth]{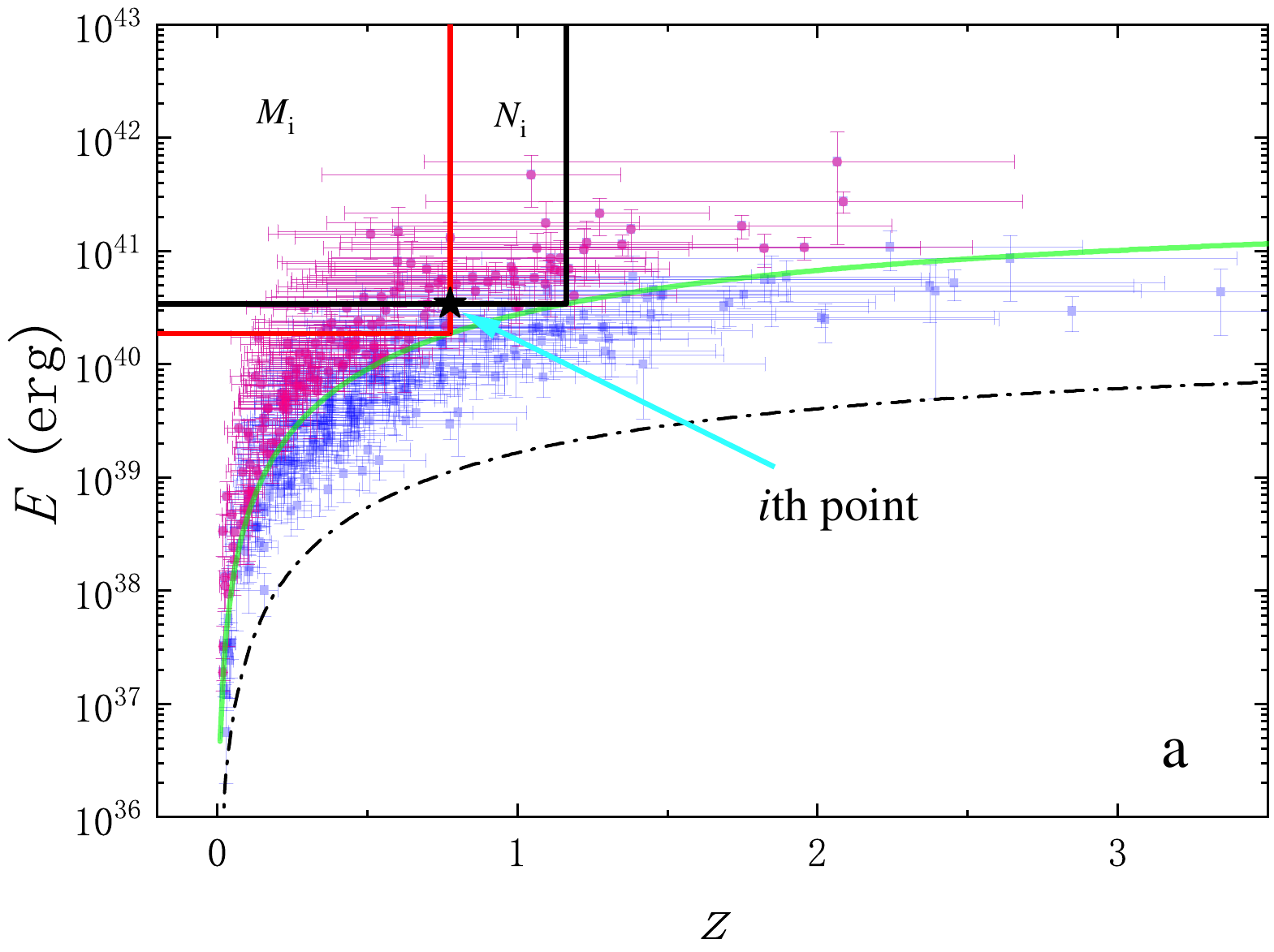}
        \includegraphics[width=0.5\textwidth]{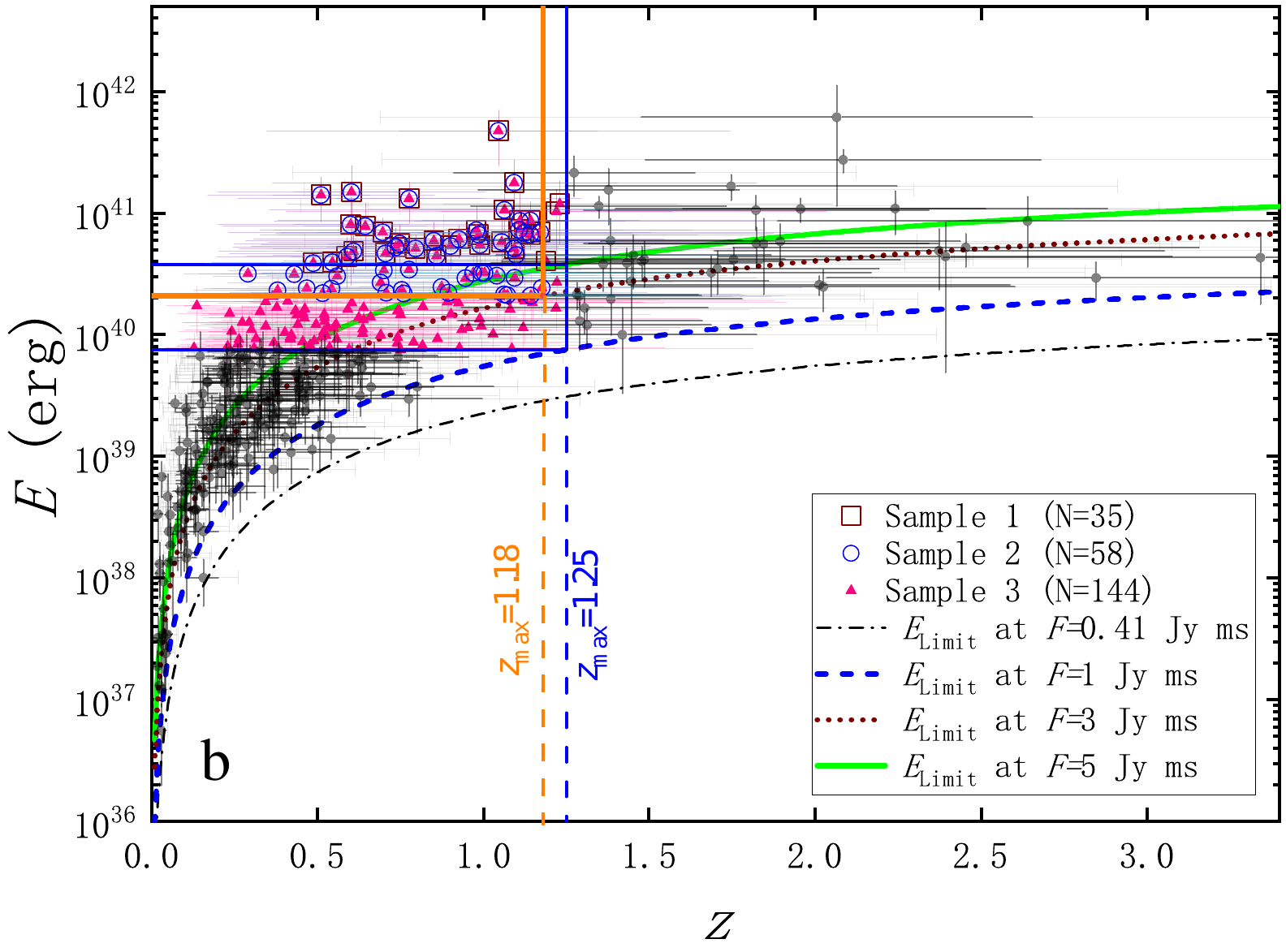}
        \caption{Energy-redshift relations of CHIME FRBs. Three volume-limited samples of 35, 58, and 144 FRBs enclosed in rectangles in panel b are drawn from the full sample of 366 FRBs above the dash-dotted line in panel a with the fluence-cuts of 5 (solid green curve), 3 (dotted purple curve), and 1 (dashed blue curve) Jy ms, correspondingly.  \label{fig:f1}}
\end{figure*}

Until March 2022, more than 700 FRBs had been observed\footnote{For a complete list of known FRBs, see \url{https://www.herta-experiment.org/frbstats} or the
Transient Name Server \citep[TNS,][]{petroff2020fast} } \citep{spanakis2021frbstats}. Notably, the CHIME/FRB Collaboration published their first catalogue of 492 FRB sources, of which 474 were non-repeaters and 18 were repeaters. 62 sub-bursts are distinguished from the 18 repeaters. To weaken the selection bias effect, we selected the FRB data detected by the CHIME telescope only,\footnote{https://www.chime-frb.ca/catalog} listed in Table 1 \citep[refer to][]{Aamiri2021first}, in which the selection criteria were that (1) FRBs with bonsai$\_$snr $<10$ were rejected, (2) FRBs with fluence detected directly after a system restarts were chosen, (3) three FRBs (FRB20190210D, FRB20190125B, and FRB20190202B) detected in far side-lobes were excluded, (4) FRBs detected during pre-commissioning, periods of low sensitivity, or on days of software upgrades were excluded, (5) the first sub-burst was taken for the repeating FRBs, and (6) the DM of intergalactic medium ($DM_{IGM}$) had to be larger than zero. As a result, 366 CHIME FRBs, including 354 non-repeating and 12 repeating sources above the minimum fluence of $0.41$ Jy ms, are included in our full sample. It is worth noting that the full sample is fluence-limited and largely affected by the instrumental effect, particularly for the fainter FRBs. To reduce the observational bias, we followed \cite{2015ApJ...806..125P} in using higher-fluence cuts of 5, 3, and 1 Jy ms to build three volume-limited samples of 35 (sample 1), 58 (sample 2), and 144 (sample 3) FRBs within ranges of $[z\leq1.25, E\geq3.78\times10^{40}\rm erg]$, $[z\leq1.18, E\geq2.10\times10^{40}\rm erg]$, and $[z\leq1.25, E\geq7.56\times10^{39}\rm erg]$, correspondingly. This has been depicted in Fig. \ref{fig:f1}, in which the fluence limit of 0.41 Jy ms is also plotted for comparison. 

Phenomenologically, FRBs can be classified into repeaters and apparent non-repeaters. However,  there are some FRBs classified as ‘one-off FRBs’ that are actually repeating ones \cite[eg.,][]{kumar2019faint,chen2022uncloaking}. Interestingly, \cite{li2021long} proposed that FRBs are better to be classified into long and short groups on basis of the time durations. It is still unknown whether different kinds of FRBs have the same origin \citep[eg.,][]{caleb2019all,Genghuang2021,2024NatAs...8..337K,2023ApJ...954...80G}. Unfortunately, there is only one repeating FRB 20151125 exhibiting ultra-wide radio pulses detected by the Big Scanning Array of the Lebedev Physical Institute \citep[BSA/LPI;][]{Fedorova2019ARep}. Categorizing FRBs into repeaters and non-repeaters may be not intrinsic but superficial owing to the sensitivity limits, the observational time coverage \citep{ai2021true}, or the possible reclassification of either repeaters or one-offs \citep{2022ApJ...939...27C}, which motivates us to combine the repeating and non-repeating CHIME FRBs into a large sample uniformly. 

\section{Methods} \label{sec:method}

\subsection{Isotropic energy}\label{Energy}
Owing to the effects of frequency drift and cosmological time dialation, the obseved FRB spectra should be corrected to the rest frame in order to calculate the intrinsic parameters accurately. The isotropic energy of a FRB is traditionally determined by $E=4 \pi D_{L}^{2}(z) F_{\rm obs}\nu_{c}k(z)/(1+z)$, where $D_{\rm L}(z)= c(1+z) \int_{0}^{z} 1/ \{H_{0}\left[(1+z)^{3} \Omega_{m}+\Omega_{\Lambda}\right]^{1 / 2}\}dz$ is the luminosity distance, $\nu_c =600\,\mathrm{MHz}$ is the central frequency of CHIME \citep{bandura2014canadian,amiri2018chime,Aamiri2021first}, and $k(z)$ represents the k-correction factor. However, there are two issues about the k-correction. One is about the spectral shape, which it is difficult to determine the reliable form of, such as a power law or a Gaussian profile \citep{pleunis2021fast}. Another one is how to extrapolate the credible bandwidth in the rest-frame. To avoid the influence of the above two issues on the calculation accuracy of energy, we followed \cite{hashimoto2022energy} to compute the isotropic energy at $\nu_c=600$ MHz within a frequency band of $400-800$ MHz via
\begin{equation}
E=E_{iso}=4 \pi D_{\rm L}^{2}(z) E_{\rm obs, 400}/(1+z),
\end{equation}
where $E_{\rm obs, 400}$ is the observed energy integrated over the rest-frame 400 MHz. It can be estimated with
\begin{equation}
        E_{\rm obs, 400}=\begin{cases}
                F_{\nu}(\frac{400\times10^6}{Hz}) & (\Delta\nu_{\rm obs,\rm itg}\geq\Delta\nu_{\rm obs,\rm FRB})\\
                F_{\nu}(\frac{400\times10^6}{Hz})(\frac{\Delta\nu_{\rm obs,\rm itg}}{\nu_{\rm obs,\rm FRB}}) & (\Delta\nu_{\rm obs,\rm itg}<\Delta\nu_{\rm obs,\rm FRB}),
                \end{cases}
        \label{E-dis}
\end{equation}
in which $\Delta\nu_{\rm obs,\rm itg}$ is defined as the integration width in the observer-frame, corresponding to an integration width of 400 MHz too in the rest frame. The energy limit at a given redshift, $z$, is approximated by $E_{\text {limit }}=4 \pi D_{\mathrm{L}}^{2}(z) E_{\rm obs,\min }/(1+z)$, with a fluence sensitivity of 5, 3, or 1 Jy ms in Eq. (\ref{E-dis}). It is necessary to emphasize that the maximum redshift, $z_{\rm max}$, or the minimum energy, $E_{\rm min}$, in Fig. \ref{fig:f1}b are self-consistently evaluated to ensure that the number of FRBs within the rectangle possesses the largest value. 

\subsection{Non-parametric method}\label{non-para}
The Lynden-Bell's $c^{-}$ method adopted in this paper has been widely used to analyze the luminosity distribution, isotropic energy distribution, and redshift distribution of quasars \citep{lynden1971method,EP1992simple}, gamma-ray bursts \citep[GRBs; see, e.g.,][]{yu2015unexpectedly,petrosian2015cosmological,liu2021isotropic,2022MNRAS.513.1078D}, and FRBs \citep[see, e.g.,][]{deng2019energy}. This method can break the degeneracy between the formation rate and the evolution of the $L$ or $E$ function, as long as the sample is large enough. If the energy function does not evolve with redshift, one can write $\Phi(E, z)$ as $\Phi(E, z)=\Psi(E) \rho(z)$, where $\Psi(E)$ is the isotropic energy function and $\rho(z)$ is the FRB formation rate at $z$. However, the energy function, $\Psi(E(z))$, usually evolves with redshift $z$ as $\Phi(E, z)=\Psi(E(z)) \rho(z)$ instead of $\Phi(E, z)=\Psi(E) \rho(z)$, and this degeneracy can be eliminated with $g(z)=(1 + z)^k$, so $\Phi(E, z)=\rho(z)\Psi(E(z) / g(z)) = \rho(z)\Psi(E_{0}) $, where $\Psi(E_0)$ is independent of redshift and represents the local energy function at $z=0$. Then, the isotropic energy at redshift $z = 0$ is $E_{0}=E(z)/g(z)$, with $E_{0}$ being independent of $z$. Observationally, the energy and the redshift of the CHIME FRBs are correlated, as is displayed in Fig. \ref{fig:f1}. This may result from instrumental effects and is not an intrinsic dependence. Thus, as the first step, we utilized the non-parametric method to reduce the observational bias and obtain the intrinsic correlations between $E$ and $z$ of samples 1, 2, and 3 in advance. 

Following some previous works \citep[eg.,][]{lynden1971method,EP1992simple,deng2019energy,liu2021isotropic,zeng2021cosmological,2022MNRAS.513.1078D}, we used the non-parametric method to derive the value of $k$ \citep{EP1992simple}. For the i\textit{th} data point $(E_i, z_i)$ in the $E-z$ plane of Fig. \ref{fig:f1}a,  we defined the set $J_i$ and $J_{i}^{\prime}$ as
\begin{equation}
\begin{aligned}
&J_{i}=\left\{j \mid E_{j} > E_{i},\,E_{j,min}(z)< E_{i}(\rm{or}\, z_{j} < z_{i,max})\right\}\\ and\\
&J_{i}^{\prime}=\left\{j \mid  z_{j}<z_{i},\,z_{j,max}> z_i (\rm{or}\, E_{j} > E_{i,min})\right\},
\end{aligned}
\end{equation}
where $z_{i,max}$ is the maximum redshift at which a FRB with the energy $E_i$ can be detected by CHIME, and the parameter $E_{i,min}$ is the minimum energy that can be detected at redshift
$z_i$. The region $J_i$ is shown in Fig. \ref{fig:f1}a as a black rectangle, and the number of FRBs contained in this region is $N_{i}$. Similarly, the number of FRBs in $J_{i}^{\prime}$ (see red rectangle in Fig. \ref{fig:f1}) is denoted as $M_i$. Independence between $E_i$ and $z_i$ would make the number of the sample,
\begin{equation}
R_{i}=\rm{number} \left\{j \in \mathit{J_{i}} \mid z_{j} \leqslant z_{i}\right\}
,\end{equation}
uniformly distribute between 1 and $N_i$ \citep{EP1992simple}. The expected mean and the variance are $E_{i}=\left(N_{i}+1\right) / 2$ and $V_{i}=\left(N_{i}-1\right)^{2} / 12$, respectively.

\begin{figure*}
        \resizebox{0.33\hsize}{!}{\includegraphics{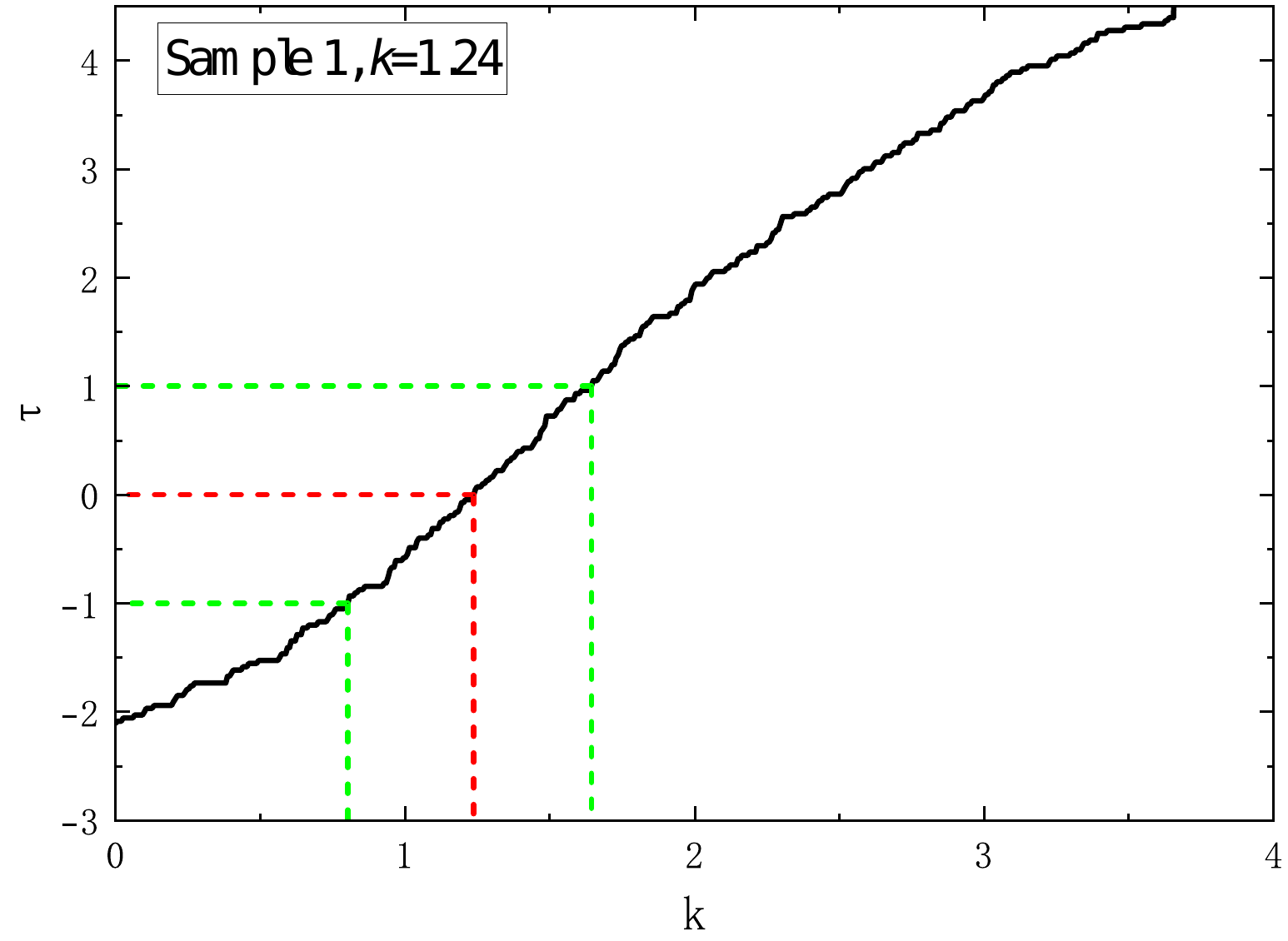}}
        \resizebox{0.33\hsize}{!}{\includegraphics{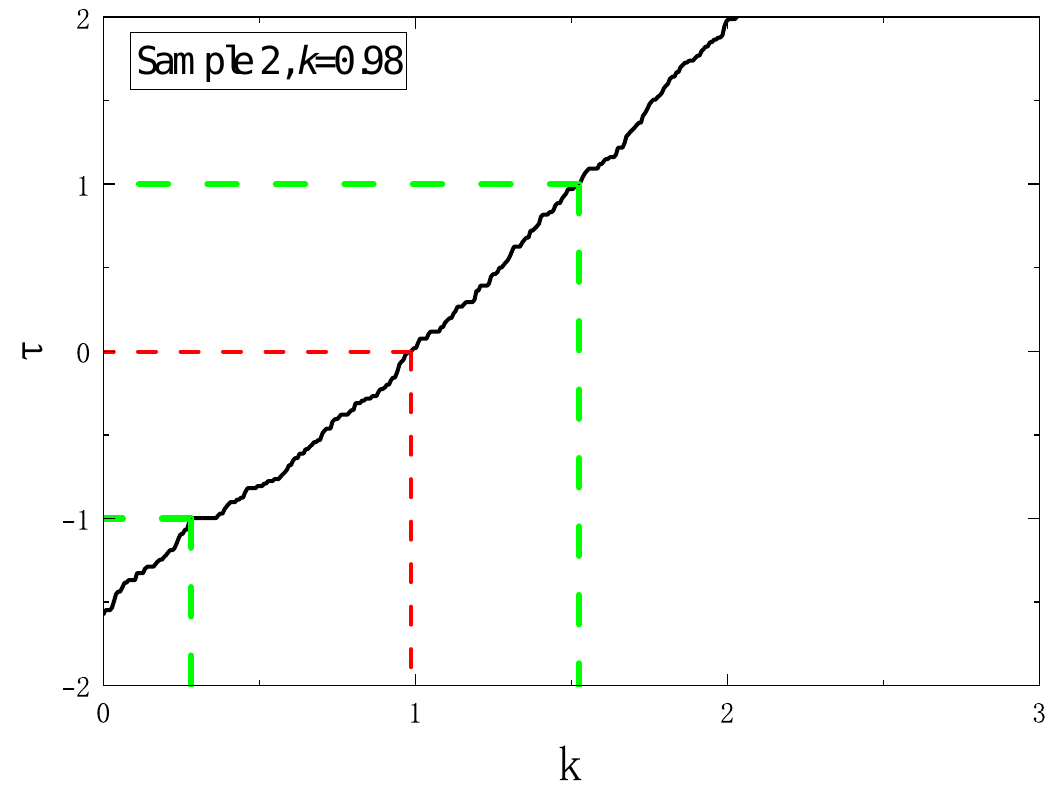}}
        \resizebox{0.33\hsize}{!}{\includegraphics{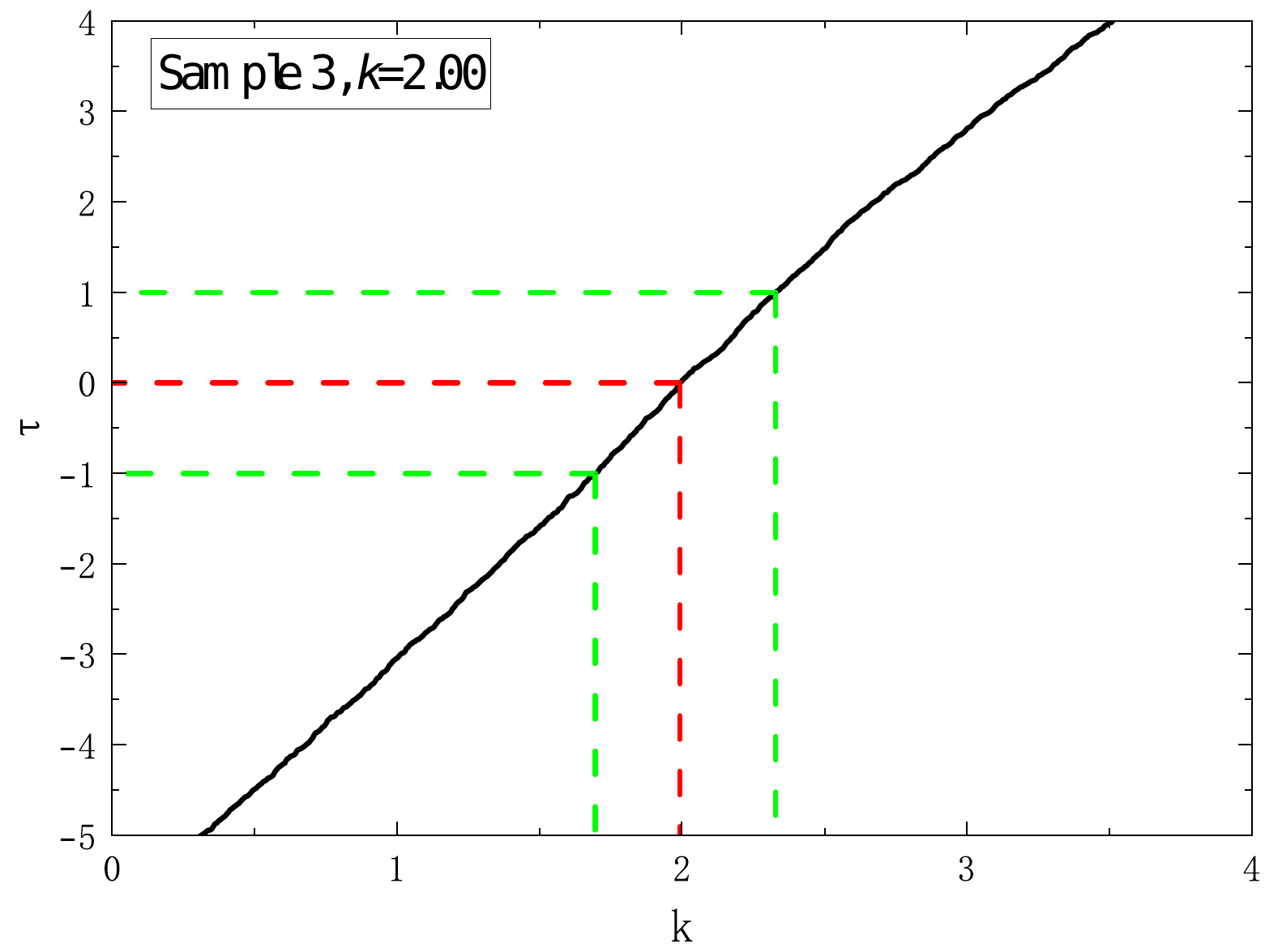}}
        \caption{$\tau$ parameter versus the power-law index, $k$, for different volume-limited  samples. The value of $\tau = 0$ (dashed red line) showing no correlation gives the best estimates of $k$ within a 1 $\sigma$ range (dashed green lines), i.e. $k=1.24^{+0.40}_{-0.44}$ for sample 1, $k=0.98^{+0.54}_{-0.69}$ for sample 2, and $k=1.99^{+0.34}_{-0.30}$ for sample 3, respectively.\label{fig:f2}}
\end{figure*}

\begin{figure*}
        \resizebox{0.33\hsize}{!}{\includegraphics{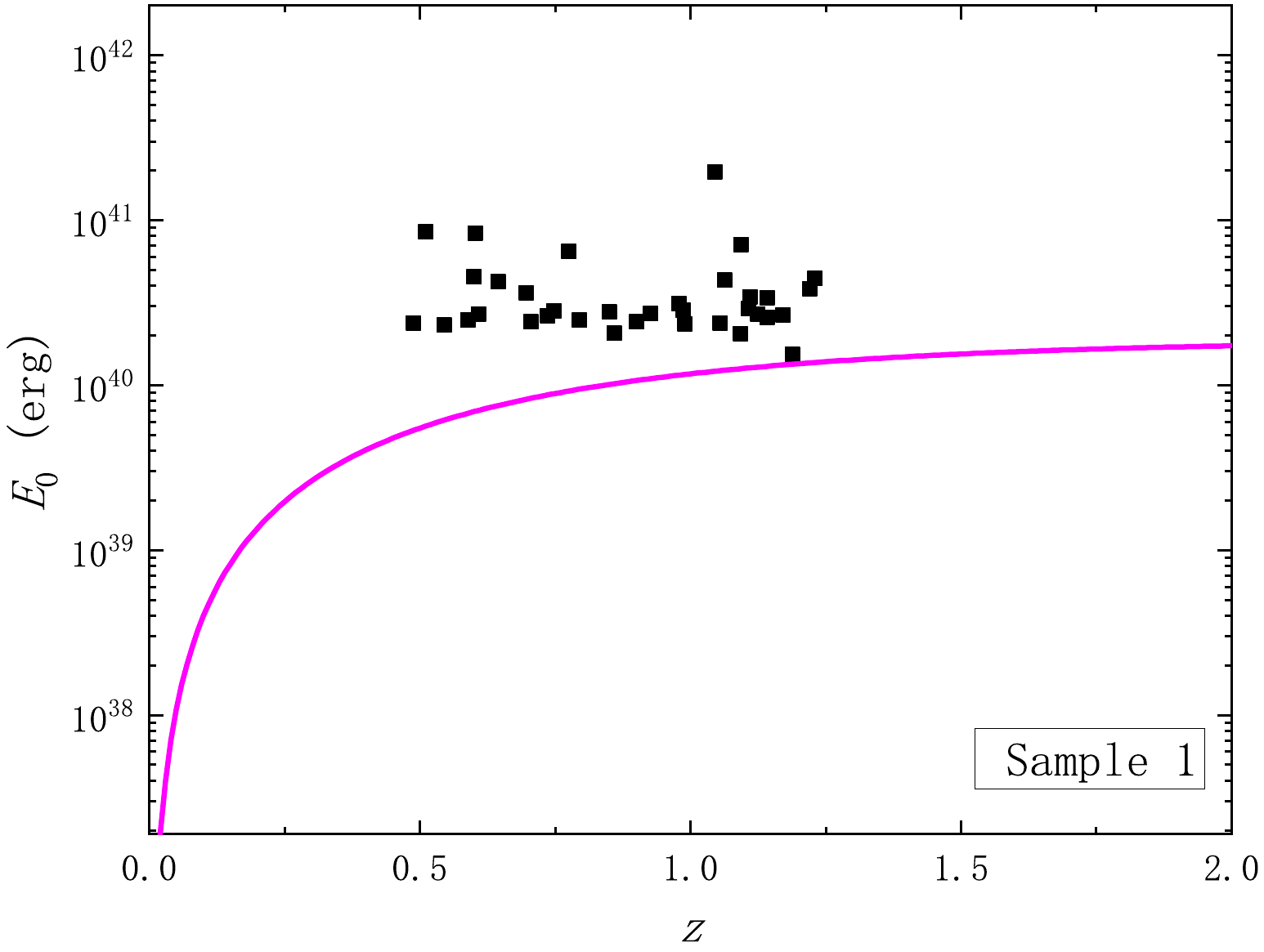}}
        \resizebox{0.33\hsize}{!}{\includegraphics{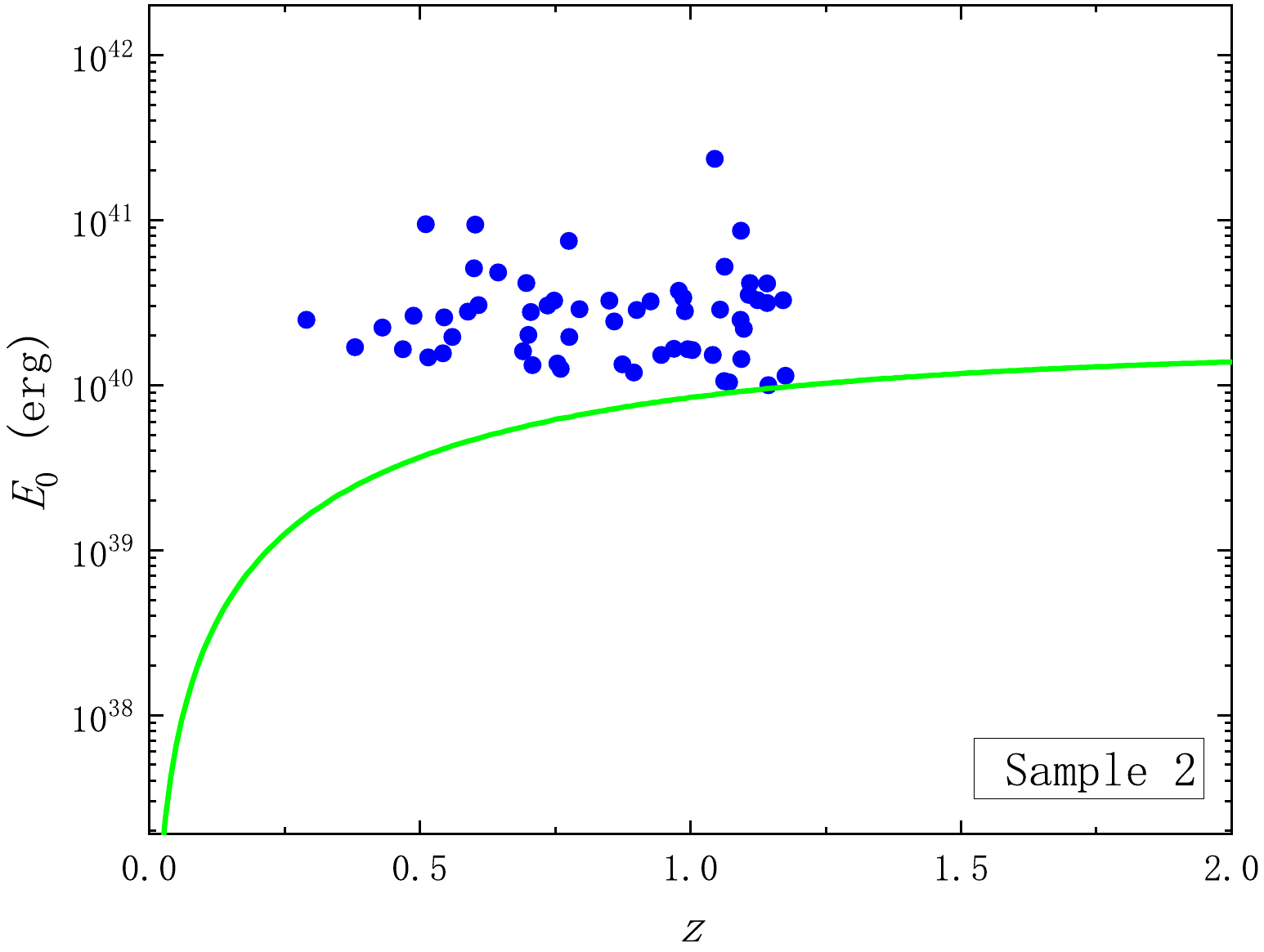}}
        \resizebox{0.33\hsize}{!}{\includegraphics{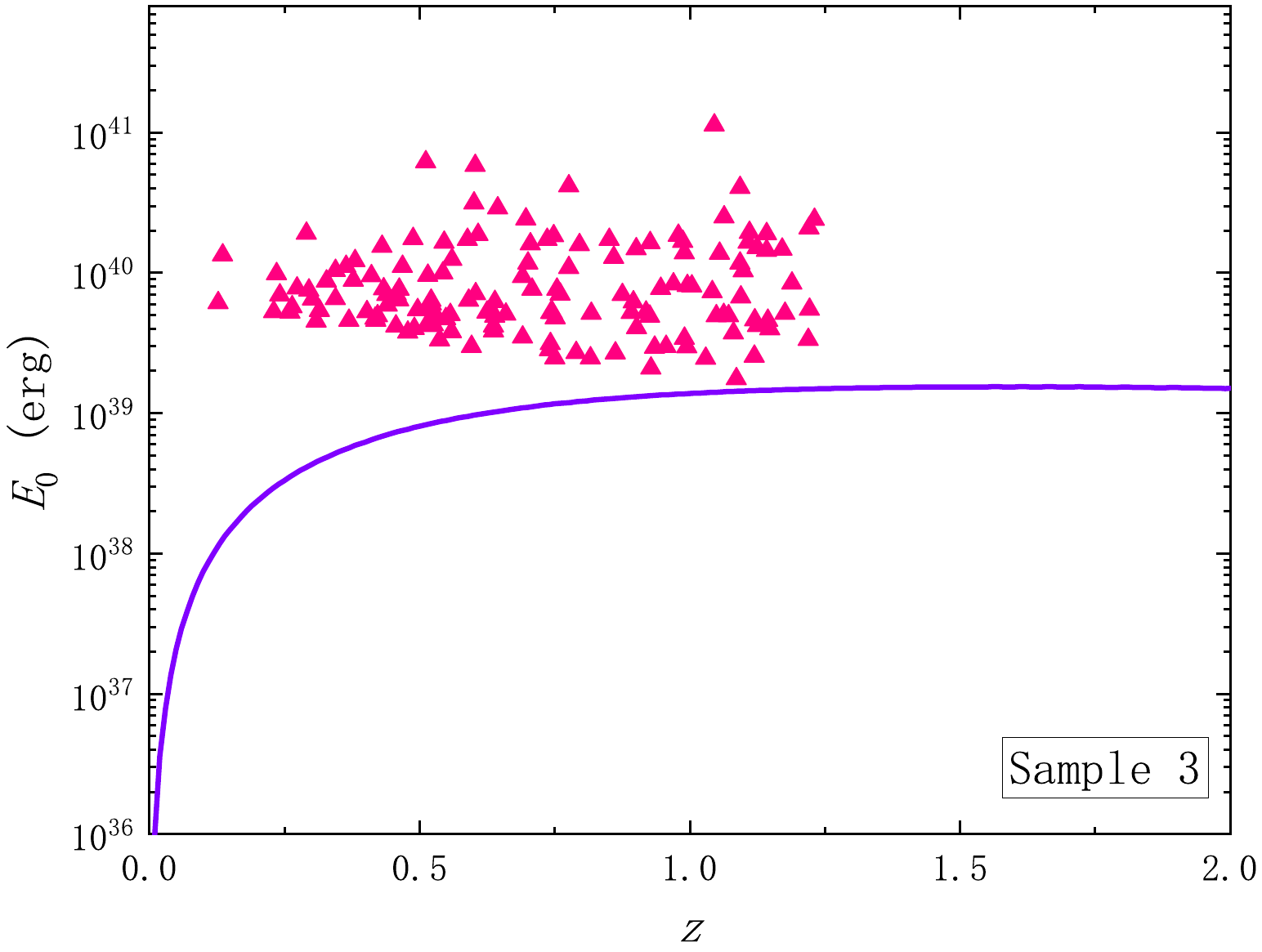}}
        \caption{De-evolved energy plotted against redshift as $E_{0} = E(z)/(1+z)^{1.24}$ for sample 1,  $E_{0} = E(z)/(1+z)^{0.98}$ for sample 2, and $E_{0} = E(z)/(1+z)^{1.99}$ for sample 3. The solid lines show the corresponding truncation line or fluence limit. \label{fig:f3}}
\end{figure*}

The statistic, $\tau$, used to test the dependence between energy and redshift is defined as
\begin{equation}
\tau \equiv \sum_{i} \frac{\left(R_{i}-E_{i}\right)}{\sqrt{V_{i}}}.
\end{equation}
If energy and redshift do not have any correlations, the $\tau$ statistic gives $\tau=0$. We can then make the transformation $E(z) \rightarrow E_{0}=E(z) / g(z)$ and vary $k$ until $\tau \rightarrow 0$, as well as until the error is reported at a $1\sigma$ confidence level (when $\tau=\pm 1$). Applying the non-parametric method to the three volume-limited samples as is shown in Fig. \ref{fig:f1}b, one can obtain the corresponding $k$ values as the $\tau$ is equal to zero. Figure \ref{fig:f2} shows the variation in the $\tau$ value with $k$ and returns $k=1.24^{+0.40}_{-0.44}$ for sample 1, $k=0.98^{+0.54}_{-0.69}$ for sample 2, and $k=1.99^{+0.34}_{-0.30}$ for sample 3, respectively, at a $1\sigma$ confidence level. Then, we apply the energy evolution form of $g(z)=(1+z)^{k}$ to remove the evolution effect. The de-evolved energy can be expressed as $E_{0}=E(z) /(1+z)^{k}$ and the distribution of $E_{0}$ and $z$ is shown in Fig. \ref{fig:f3}.

\section{The energy function evolved or de-evolved with redshift}\label{energyfun}\label{Evsz}
Through $M_i$ and $N_i$, depicted in Fig. \ref{fig:f1}a, we next derived the de-evolved energy function, $\Psi(E_0)$, and the cumulative redshift distribution,
$\Phi(z)$, with the following non-parametric method \citep{lynden1971method,EP1992simple}, 
\begin{equation}
\Psi\left(E_{0,\mathrm{i} }\right)=\prod_{j<i}\left(1+\frac{1}{N_{\mathrm{j}}}\right),
\label{eq:EF}
\end{equation}
and
\begin{equation}
\phi\left(z_{\mathrm{i}}\right)=\prod_{j<i}\left(1+\frac{1}{M_{\mathrm{j}}}\right),
\label{eq:ZF}
\end{equation}
where $ j < i$  means that the energy, $E_{0,j}$, is greater than $E_{0,i}$ in Eq.~(\ref{eq:EF}) and the FRB has a redshift, $z_j$, less than $z_i$ in Eq.~(\ref{eq:ZF}). Subsequently, the FRB formation rate $\rho(z)$ can be determined with the following formula:
\begin{equation}
\rho(z)=\frac{d \phi(z)}{d z}(1+z)\left(\frac{d V(z)}{d z}\right)^{-1},
\label{eq:FR}
\end{equation}
where the factor (1 + z) denotes the correction for the effect of cosmic time dilation. The differential co-moving volume, $dV(z)/dz$, was taken in the following form \citep{yu2015unexpectedly,qiang2022fast}:
\begin{equation}
\frac{d V(z)}{d z}=\frac{c}{H_{0}} \frac{4 \pi D_{\mathrm{L}}^{2}}{(1+z)^{2} \sqrt{\Omega_{m}(1+z)^{3}+\Omega_{\Lambda}}},
\label{eq:dV/dz}
\end{equation}
for a flat $\Lambda$CDM model.

\begin{figure*}
        \centering
        \resizebox{0.33\hsize}{!}{\includegraphics{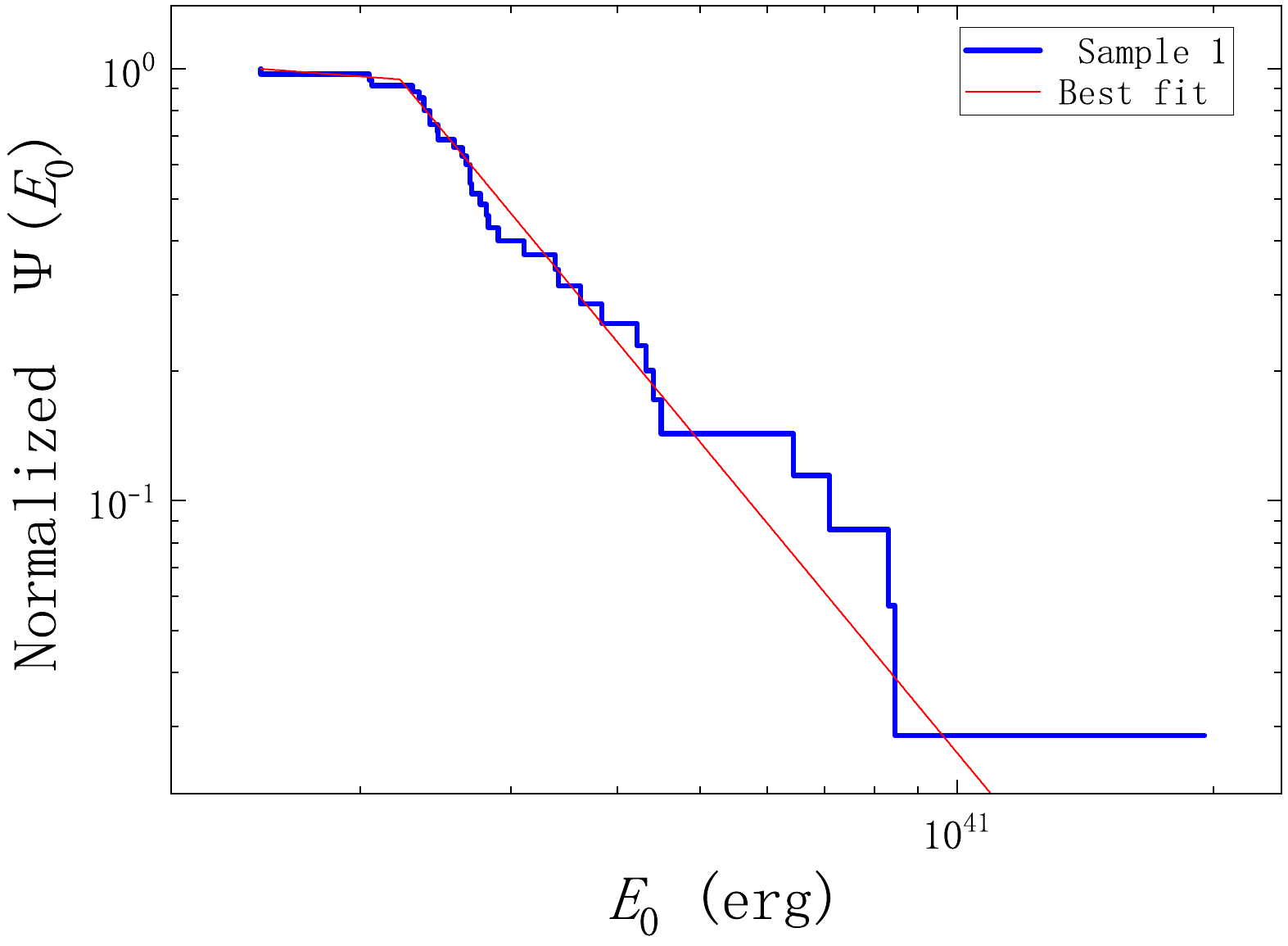}}
        \resizebox{0.33\hsize}{!}{\includegraphics{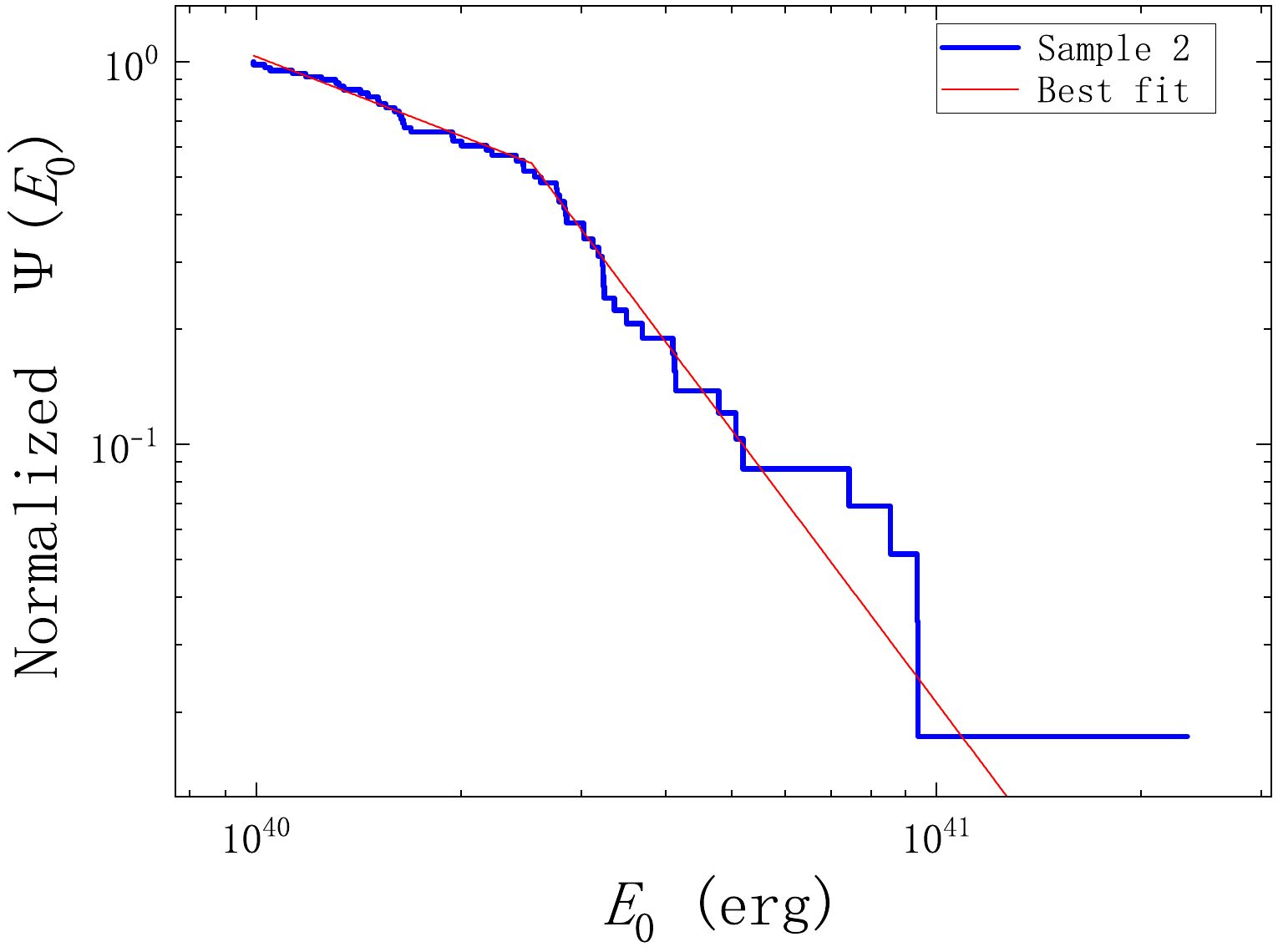}}
        \resizebox{0.33\hsize}{!}{\includegraphics{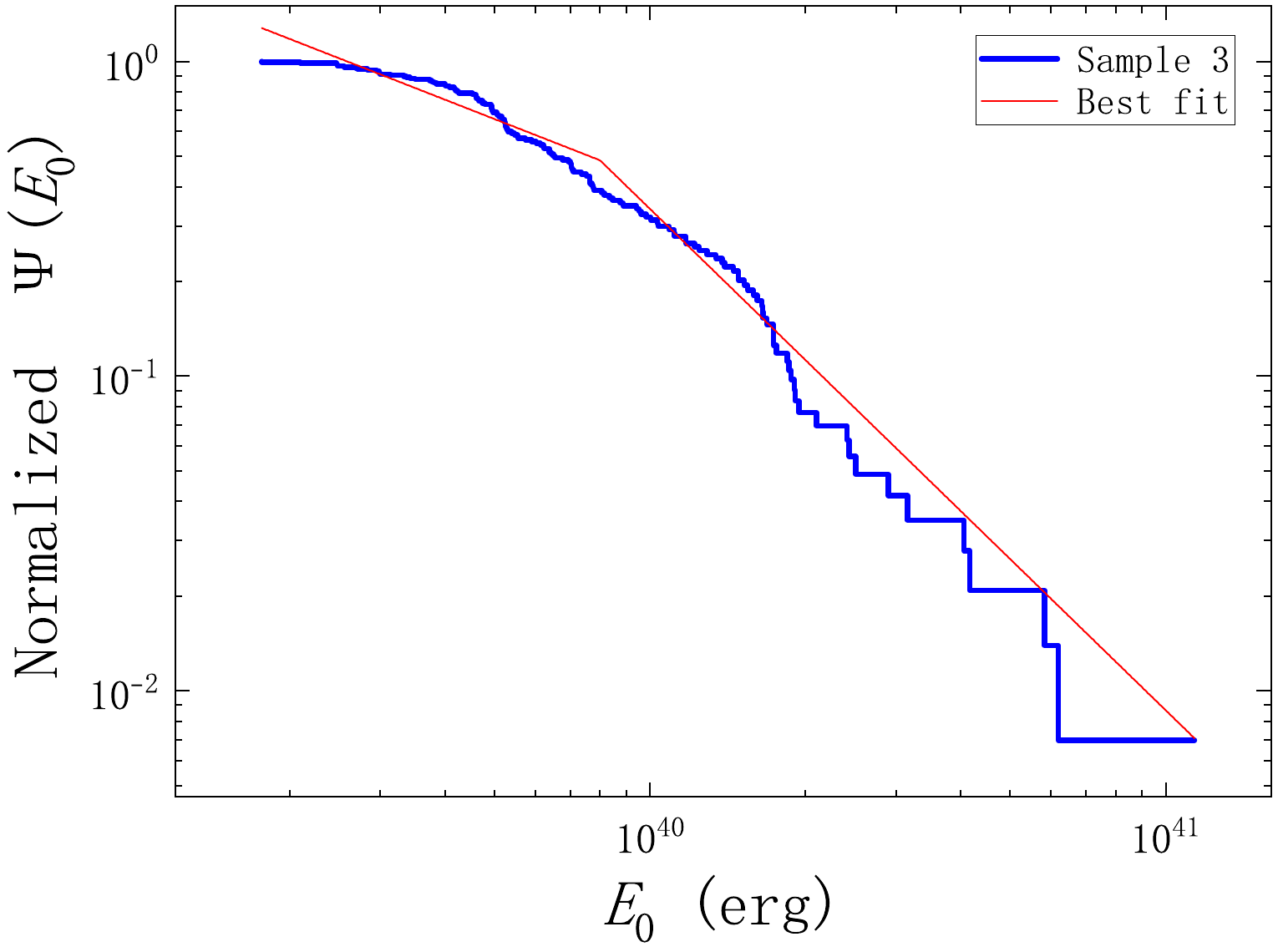}}
        \caption{Energy function of FRBs. The blue survival lines of $\Psi(E_{0})$ for the three FRB samples are normalized to unity at the lowest energy. The best fits with a BPL form are symbolized with broken red lines. The meaning of the symbols is shown in the insert. \label{fig:f4}}
\end{figure*}

Using the new data set ($N_{\mathrm{j}}$), one can obtain the normalized cumulative energy function, $\Psi(E_{0})$, in Fig. \ref{fig:f4}. For sample 1, the de-evolved energy function can be described by a broken power-law (BPL) function as
\begin{equation}
        \Psi(E_{0}) \propto \begin{cases}
                E_{0}^{-0.15 \pm 0.16}, & E_{0}<E_{b}\\
                E_{0}^{-2.39 \pm 0.10}, & E_{0} \geq E_{b}\end{cases}, 
        \label{E-dis-sample1}
\end{equation}
where the broken energy is about $E_{b} = 2.22\times 10^{40}$ erg. For sample 2, we have 
\begin{equation}
        \Psi(E_{0}) \propto \begin{cases}
                E_{0}^{-0.69 \pm 0.02}, & E_{0}<E_{b}\\
                E_{0}^{-2.37 \pm 0.11}, & E_{0} \geq E_{b}\end{cases},
        \label{E-dis-sample2}
\end{equation}
where the broken energy is about $E_{b} = 2.54 \times 10^{40}$ erg. For sample 3, we obtain
\begin{equation}
        \Psi(E_{0}) \propto \begin{cases}
                E_{0}^{-0.64 \pm 0.02}, & E_{0}<E_{b}\\
                E_{0}^{-1.59 \pm 0.07}, & E_{0} \geq E_{b}\end{cases},
        \label{E-dis-sample3}
\end{equation}
where the broken energy is about $E_{b}=8.00 \times 10^{39}$ erg. The broken energies are roughly consistent with some previous results \citep[e.g.,][]{deng2019energy}. It is worth noting that the energy function only represents the `local' form at $z = 0$. The energy function at a given redshift, $z$, should be $\Psi(E_{0})(1+z)^{k}$ in the observer frame and roughly break at $E_{b}(1+z)^{k}$ .
\section{The redshift distribution and formation rate evolution } \label{sec:zdisandFm}

\begin{figure}
\centering 
\includegraphics[width=0.5\textwidth]{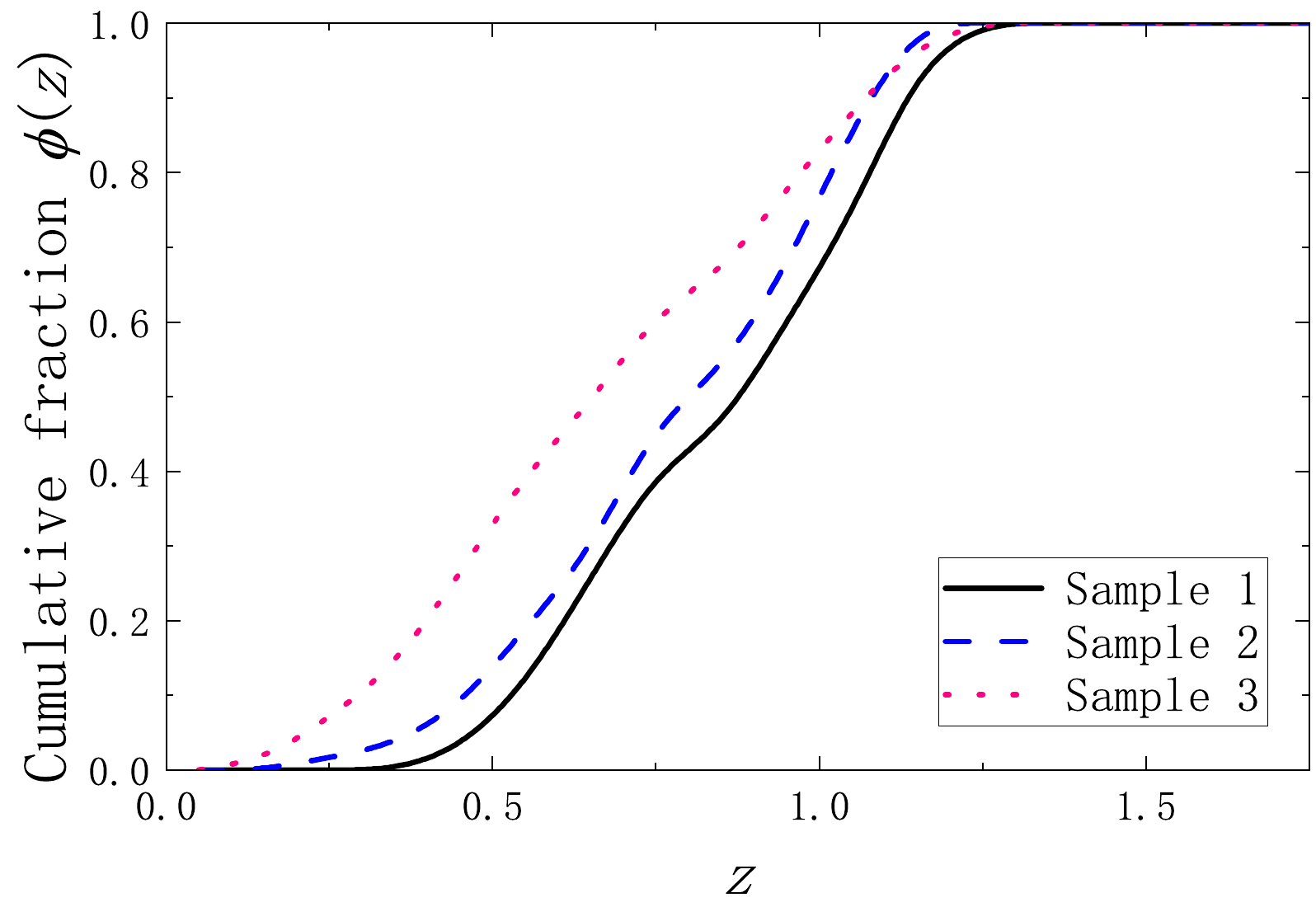}
\caption{Cumulative redshift distributions of sample 1 (solid line), sample 2 (dashed line), and sample 3 (dotted line).\label{fig:f5}}
\end{figure}

Figure \ref{fig:f5} shows the  normalized cumulative redshift distribution. A Kolmogorov-Smirnov (K-S) test gives the statistic $D=0.12$ (<$D_\alpha=0.28$ for $\alpha=0.05$), with a $p$ value of 0.87, indicating that the redshift distributions of samples 1 and 2 are identical, while the K-S test between samples 1 and 3 returns 
$D=0.29$ (>$D_\alpha=0.25$ for $\alpha=0.05$), with $p=0.01$, showing that the redshift distributions of both samples are taken from different parent distributions. Similarly, we obtain $D=0.22$ (>$D_\alpha=0.21$ for $\alpha=0.05$) and $p=0.03$ between samples 2 and 3, which indicates the two samples are differently distributed.

\begin{figure}
                        \includegraphics[width=0.5\textwidth]{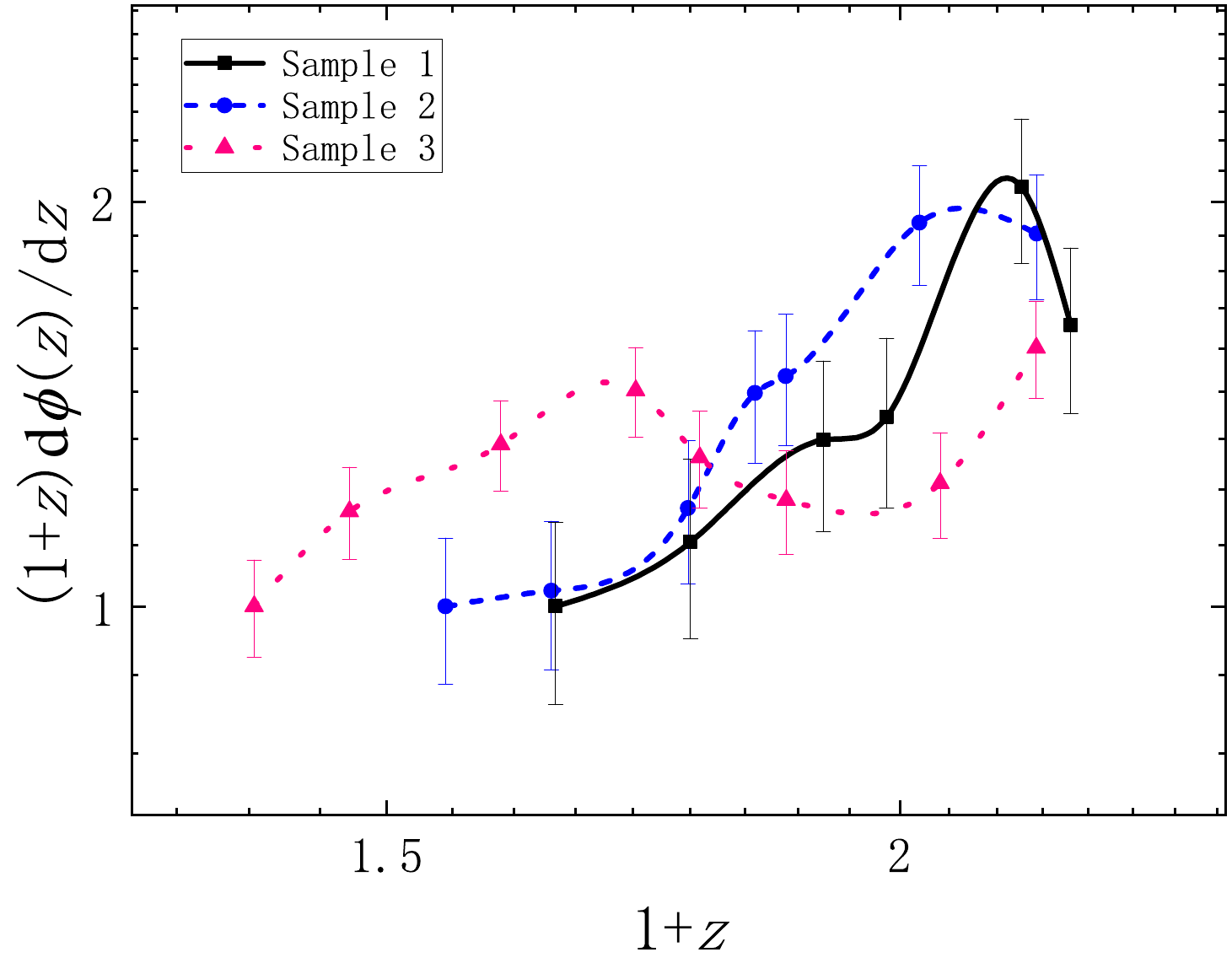}
        \caption{Evolution of $(1+z) d\phi(z)/dz$ with $z$ for sample 1 (filled squares), sample 2 (filled circles), and sample 3 (filled triangles). The data have been normalized to unity at the first redshift point.\label{fig:f6}}
\end{figure}

\begin{figure}
\centering
\includegraphics[width=0.5\textwidth]{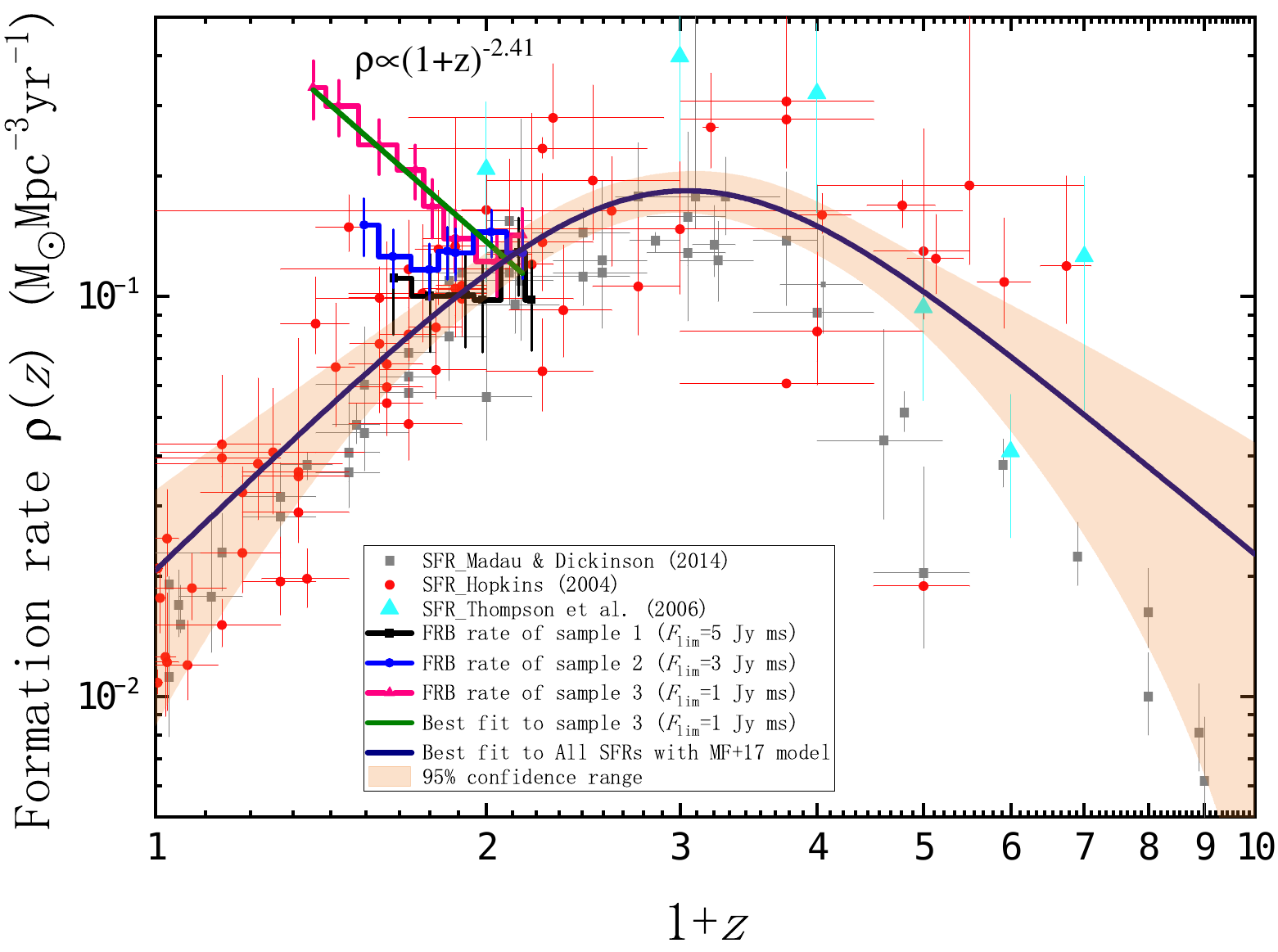}
\caption{Formation rate, $\rho(z)$, of CHIME FRBs and the SFR versus redshift. The thick blue curve and the shaded regions show, respectively, the best fit of $SFR=0.02(1+z)^{2.65}/[1+((1+z)/2.95)^{4.93}]$ with the theoretical model in \cite{madau2017radiation} and a 2$\sigma$ confidence level. Grey squares, blue upward triangles, and red circles represent the observed SFR taken from \citet{madau2014cosmic}, \citet{thompson2006star} and \citet{hopkins2004evolution}, respectively. The three thick step lines denote the FRB event rates of the three samples. The best fit to sample 3 is displayed by the solid green line.\label{fig:f7}}
\end{figure}

We plot the differential distribution, $d \phi(z)/dz$, of the redshift for the above three FRB samples in Fig. \ref{fig:f6}, in which it is interestingly found that $(1+z)d\phi(z) / dz$ of samples 1 and 2 exhibit a very similar evolution with $z$. They increase gradually at lower redshifts and then quickly decrease at higher redshifts of $z\geq1$. However, sample 3 evolves with redshift in a different way, especially at the lower redshift range. Next, we used Eq.~(\ref{eq:FR}) to estimate the FRB formation rate and plot its relation with redshift in Fig. \ref{fig:f7}. It can clearly be found that the event rate of the FRBs in samples 1 and 2 roughly coincides with the SFR, despite a slight discrepancy at the lower-redshift end for sample 2. Surprisingly, the formation rate of sample 3 exceeds the star formation rate (SFR) at the lower redshift of $z\leq1$, which is consistent with the results of GRBs discovered in some previous works \cite[e.g.][]{2022MNRAS.513.1078D,2023ApJ...958...37D}. Observationally, the formation rate of the FRBs in sample 3 can be simply fitted by a single power-law form as
\begin{equation}
        \rho(z)\varpropto(1+z)^{-2.41\pm0.40}\ \ (\rm {M_\odot\  Mpc^{-3}\ yr^{-1}}).      
        \label{eq:ratefiting}
\end{equation}
The higher event rate of sample 3 FRBs at lower redshifts demonstrates that most of the FRBs in this sample probably originated in older stellar popuplations in the nearby Universe. Furthermore, the consistency of the event rate declining with redshift for the three FRB samples suggests that the expected numbers of faint FRBs at higher redshifts below the threshold are too small to influence the total parameter features and the FRB rate significantly. It is curious to note that FRB 20121102A also follows the degressive trend, when its characteristic energy, redshift, and SFR are considered together \citep{2021Natur.598..267L}.

\begin{figure}
\centering
\includegraphics[width=0.5\textwidth]{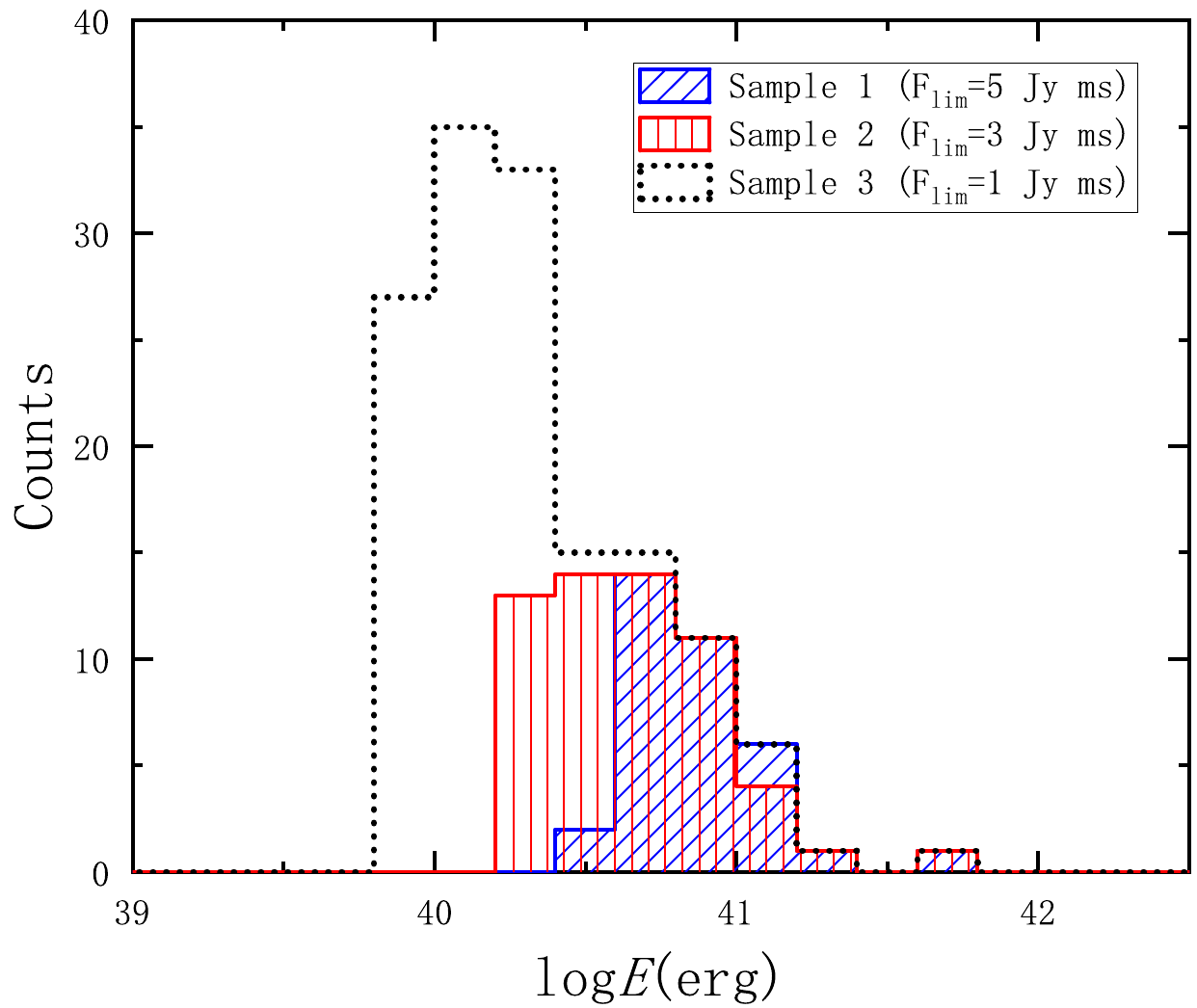}
\caption{Histograms of isotropic energy of FRBs in samples 1, 2, and 3. The average energies become larger as the fluence limits increase.  \label{fig:f8}}
\end{figure}

In practice, there is some evidence that FRBs might follow the global SFR of the Universe, which is confirmed by our samples 1 and 2. For example, FRB 20180916B and FRB 20201124A are detected from the star-forming regions \citep{marcote2020repeating,piro2021fast}.  Simultaneously, there is more evidence showing that at least parts of FRBs originate in magnetars \citep{bochenek2020fast,andersen2020bright,heintz2020host,li2020comparative,mereghetti2020integral,bochenek2021localized}, which means that the formation rate of such FRBs would follow the cosmic SFR \citep{Aamiri2021first,zhang2021energy,james2022fast}, similar to the samples of Parkes and ASKAP FRBs \citep{deng2019energy,zhang2021energy,james2022fast}. Noticeably, parts of previous results may be somewhat biased by the limited FRB sample sizes. 

It is worth emphasizing that our samples 2 and 3 are already large enough and the non-parametric method used in this work does not rely on any prior assumptions. Our results support the recent finding that the CHIME FRB population does not track the SFR \citep{hashimoto2022energy,qiang2022fast,zhang2022chime}. A significantly delayed FRB distribution tracking the star formation is also found to be reasonable \citep[see, e.g.,][]{CaoXF2018FRBs,zhang2021energy, qiang2022fast,zevin2022observational,panther2023most,2023ApJ...954...80G}. This indicates that FRBs could be produced from multiple progenitor channels \citep[e.g.,][]{zhang2022chime,2023ApJ...954...80G} or originate in distinct sub-populations \citep[e.g.,][]{good2023nondetection,michilli2023subarcminute,nimmo2023burst,2024SciBu..69.1020Z}, such as repeaters versus non-repeaters \citep{petroff2022fast}, or long versus short FRBs \citep{li2021long}. However, our recent investigations based on the intensity function show that the repeating and non-repeating FRBs have similar event rates on the whole sky, hinting that both types of FRBs could share the same physical origin \citep{LiLB2017,ZhangKJ2022}. Of course, the possibility that the FRB rate exceeds the SFR at lower redshifts owing to the selection function cannot be fully ruled out. Figure \ref{fig:f8} shows that the energy distributions of samples 1 and 2 are identical and they are clearly different from sample 3. On average, the isotropic energies of sample 3 FRBs are smaller than the ones of the other two samples. This may hint that the high-energy FRBs are produced from the active star forming regions, while the low-energy FRBs unassociated with the SFR are mainly formed from the quiescent regions dominated by old star populations such as compact binary mergers, which is similar to the results for long GRBs derived by \cite{2023ApJ...958...37D}.

\section{The local event rates of CHIME FRBs}\label{sec:localeventrate}
According to \cite{deng2019energy} , the number of CHIME FRBs can be determined with
\begin{equation}
N=\frac{\Omega T}{4\pi}\int_{z_1}^{z_2} \mathrm{d}z\int_{E_{min}}^{E_{max}}\frac{\Phi (E,z)}{1+z}\frac{\mathrm{d}V}{\mathrm{d}z}\mathrm{d}E,
\end{equation}
where $z_1$ and $z_2$ are the minimum and maximum redshifts, respectively, and $E_{min}$ and $E_{max}$ represent the lowest and highest energies of FRBs in the sample. $\Phi (E,z)$ is the total energy function defined in Sect. \ref{sec:method}. When the variables $E$ and $z$ are separated, we can utilize $\Phi (E,z)=\Psi(E_0)\rho(z)$ and Eqs. (\ref{E-dis-sample1}---\ref{E-dis-sample3}) to rewrite the FRB number as

\begin{equation}
        N=\frac{\Omega T}{4\pi}\int_{z_1}^{z_2} \mathrm{d}z\int_{E_{0,min}}^{E_{0,max}}\frac{\Psi(E_0) \rho(z)}{1+z}\frac{\mathrm{d}V}{\mathrm{d}z}\mathrm{d}E_{0},
\end{equation}
in which $\rho (z)$ and $\mathrm{d}V/\mathrm{d}z$ have been individually given in Eqs. (\ref{eq:FR}) and (\ref{eq:dV/dz}). The event rates of FRBs at a given redshift, $z$, can be expressed by $\rho(z)\varpropto \rho_{1}(0)$ for sample 1, $\rho(z)\varpropto \rho_{2}(0)$ for sample 2, and $\rho(z)\varpropto \rho_{3}(0)(1+z)^{-2.41}$ for sample 3, where $\rho (0)$ denotes the local event rate of FRBs at a redshift of $z=0$. Here, we adopted the CHIME's effective observation period of 214.8 days \citep{Aamiri2021first}, a field of view (FoV) of 200 $\rm {deg^2}$ \citep{amiri2018chime}, and the total number of each volume-limited sample to estimate the local FRB rate as $\rho_1(0) \approx2.06\times 10^4 \rm{\,Gpc^{-3}yr^{-1}}$, $\rho_2(0) \approx1.93\times 10^4 \rm{\,Gpc^{-3}yr^{-1}}$, and $\rho_3(0) \approx7.42\times 10^3 \rm{\,Gpc^{-3}yr^{-1}}$, rates that are roughly consistent with the ones of Parkes FRBs \citep{CaoXF2018FRBs} but slightly larger than the ones of ASKAP FRBs \citep{deng2019energy}. Notably, our results for high-energy FRBs in samples 1 and 2 are approximately consistent with the local event rate of high-energy estimated by \cite{ZhangZL2023FRBrate}. The discrepancies could be caused by the instrumental biases and some potential sample selection effects. Interestingly, it was shown by simulations that the local rate of WD mergers can be a few $10^4 \rm{\,Gpc^{-3}yr^{-1}}$, which is in good agreement with the observed SN Ia rate \citep{LiWD2011SNIa-rate}. One can also convert the local CHIME FRB rates of our samples to the full sky rates of $\rho_1(0) \approx 672\times2.06\times10^4\div365\approx3.79\times10^{4}\rm{\,sky^{-1} day^{-1}}$, $\rho_2(0) \approx 575\times1.93\times10^4\div365\approx3.04\times10^{4}\rm{\,sky^{-1} day^{-1}}$, and  $\rho_3(0) \approx 672\times7.42\times10^3\div365\approx1.37\times10^{4}\rm{\,sky^{-1} day^{-1}}$. They are very close to some previous estimates \cite[see e.g.][]{2013Sci...341...53T,2014ApJ...790..101S,2015ApJ...807...16L,LiLB2017,ZhangKJ2022}, but obviously larger than the general value of $\sim820\rm{\,sky^{-1} day^{-1}}$ reported by \cite{Aamiri2021first}, which may be biased by the larger fraction of low-redshift FRBs or the non-uniformity of CHIME FRB samples. It should be noted that our estimates of the local formation rate are better regarded as upper limits, since the faint high-redshift FRBs below the threshold are really missed, in contrast to the assumption that all FRBs within the CHIME FoV are detectable.

\section{Conclusions} \label{sec:concl}

We have applied the non-parametric method to investigate the energy function and formation rate of three volume-limited samples of CHIME FRBs. Our results are model-independent and do not suffer from any instrumental difference. The following conclusions can be drawn:
\begin{itemize}
  \item We find that the rest-frame energies  of CHIME FRBs evolve with redshifts as $ E(z)\varpropto(1+z)^{1.24}$ for sample 1, $E(z)\varpropto(1+z)^{0.98}$ for sample 2, and $E(z)\varpropto(1+z)^{1.99}$ for sample 3 with the Kendall's $\tau$ statistic method.
  \item After removing the evolution of energy with redshift, we derived the isotropic energy ($E_0$) function that can be well fitted by a BPL form with broken energies of $E_{b}\approx2.2\times10^{40}$ erg for sample 1, $E_{b}\approx2.5\times10^{40}$ erg for sample 2, and $E_{b}\approx8.0\times10^{39}$ erg for sample 3.
  \item The cumulative redshift distributions of samples 1 and 2 are identical. Sample 3 has a smaller median redshift and its redshift distribution is different according to the K-S test.
  \item The FRB event rates of samples 1 and 2 are found to be almost constant, while the event rate of sample 3 decreases with redshift and a single power-law form as $ \rho(z)\varpropto(1+z)^{-2.41}$, which indicates that the low-redshift excess of FRB rate is mainly attributed to the lower-energy FRBs, together with the selection function.
  \item The local FRB rates of the three FRB samples are found to have a similar value of $\sim10^4\rm{\,Gpc^{-3}yr^{-1}}$ or $\sim10^4\rm{\,sky^{-1} day^{-1}}$, which is roughly consistent with previous results.
  \item We find that the FRB formation rates of  FRB samples 1 and 2 follow the SFR well, while sample 3 does not match the SFR at $z<1$, which hints that a significant fraction of FRBs in sample 3 probably hail from older stellar populations in the Universe.
  
\end{itemize}
\section*{Data availability}
Tables 1 is only available in electronic form at the CDS via anonymous ftp to cdsarc.u-strasbg.fr (130.79.128.5) or via http://cdsweb.u-strasbg.fr/cgi-bin/qcat?J/A+A/.

\begin{acknowledgements}
We would like to thank the anonymous referee for valuable 
comments and suggestions that lead to an overall improvement of this study. This work was supported in part by National Natural Science Foundation of China
(grant Nos. 11988101, U2031118, 12041306, 12233002), the National Key R\&D
Program of China (2021YFA0718500), the National SKA Program of China No. 2020SKA0120300 and the
Natural Science Foundations (ZR2018MA030, XKJJC201901). YFH acknowledges the support from the Xinjiang Tianchi Program. We also acknowledge the usage of the archive data in the First CHIME/FRB Catalog (\url{https://www.chime-frb.ca/catalog}).
\end{acknowledgements}
\bibliographystyle{aa}
\bibliography{reference}
\end{document}